\documentclass[aps,preprint,superscriptaddress,showpacs]{revtex4}

\usepackage{graphicx}
\usepackage{float}
\usepackage{floatflt}
\DeclareGraphicsExtensions{.jpg,.pdf,.png,.eps}

\begin{document}

\title{Highly sensitive thermal conductivity measurements of suspended membranes (SiN and diamond) using a 3$\omega$-V$\ddot{o}$lklein method}

\author{A. Sikora}
\affiliation{Institut N\'EEL, CNRS-UJF, 25 avenue des Martyrs, 38042 Grenoble Cedex 9, France}

\author{H. Ftouni}
\affiliation{Institut N\'EEL, CNRS-UJF, 25 avenue des Martyrs, 38042 Grenoble Cedex 9, France}

\author{J. Richard}
\affiliation{Institut N\'EEL, CNRS-UJF, 25 avenue des Martyrs, 38042 Grenoble Cedex 9, France}

\author{C. H\'ebert}
\affiliation{Institut N\'EEL, CNRS-UJF, 25 avenue des Martyrs, 38042 Grenoble Cedex 9, France}

\author{D. Eon}
\affiliation{Institut N\'EEL, CNRS-UJF, 25 avenue des Martyrs, 38042 Grenoble Cedex 9, France}

\author{F. Omn\`es}
\affiliation{Institut N\'EEL, CNRS-UJF, 25 avenue des Martyrs, 38042 Grenoble Cedex 9, France}

\author{O. Bourgeois} \email{olivier.bourgeois@grenoble.cnrs.fr}
\affiliation{Institut N\'EEL, CNRS-UJF, 25 avenue des Martyrs, 38042 Grenoble Cedex 9, France}

\date{\today}

\begin{abstract}
A suspended system for measuring the thermal properties of membranes is presented. The sensitive thermal measurement is based on the 3$\omega$ dynamic method coupled to a V$\ddot{o}$lklein geometry. The device obtained using micro-machining processes allows the measurement of the in-plane thermal conductivity of a membrane with a sensitivity of less than 10nW/K (+/-$5\times10^{-3}$Wm$^{-1}K^{-1}$ at room temperature) and a very high resolution ($\Delta K/K =10^{-3}$). A transducer (heater/thermometer) centered on the membrane is used to create an oscillation of the heat flux and to measure the temperature oscillation at the third harmonic using a Wheatstone bridge set-up. Power as low as 0.1nanoWatt has been measured at room temperature. The method has been applied to measure thermal properties of low stress silicon nitride and polycrystalline diamond membranes with thickness ranging from 100~nm to 400~nm. The thermal conductivity measured on the polycrystalline diamond membrane support a significant grain size effect on the thermal transport.
\end{abstract}

\pacs{68.55.-a,73.61.-r,81.05.Uw,81.15.Fg}

\maketitle

\section{Introduction}

In recent years, the constant increase of interest for nanomaterials in thermal physics (nanophononics) \cite{Cahill2003,mingoNat,Lishi} and for thermoelectrics \cite{shakouriTAP,shakouriReview,hochbaum} has motivated the development of very fine thermal measurement techniques dedicated to small size systems. For instance the possibility of playing on the geometry of sample or on the nanostructuration of thin films to reduce the contribution of phonon to the thermal conductivity is a very active subject of research \cite{donadio,PRBHeron,mingoNat}; the reduced size of the studied objects requires specific experimental methods. Size effects are studied on the heat conduction (effect on the phonon mean free path \cite{chen2011}, on the dispersion relation \cite{PRBMingo2003} or on the transmission coefficient \cite{volz,donadio}) and to mention only few in phononic crystals, nanoparticles embedded in a matrix, nanomembranes or in the presence of rough surfaces etc... Moreover especially for grown thin films, the thermal properties may strongly depend along which axis they are measured and therefore their measurements need adapted experimental techniques from high temperature \cite{Cahill2003} to low temperature \cite{TopApplBourg}.

Since two decades, numerous static or dynamic methods have been developed to measure the thermal conductivity of new materials like 3$\omega$ \cite{Cahill,Dames,Duquesne}, hot wire (V$\ddot{o}$lklein method) \cite{Volk,Jacquot}, thermoreflectance \cite{Cahill2003,Ezzahri} or steady state methods \cite{shi2003,hippal,zink2004,zink2009}. The need for precise measurement of very thin films or membranes imposes to work with suspended systems. However, very few techniques permits such achievement on membrane and nanowire \cite{Jain,Volk,LuRSI,JAPBourg}, especially when a high sensitivity is necessary. Here we report on a very sensitive dynamic method based on a mix of the V$\ddot{o}$lklein and the 3$\omega$ methods to measure the in-plane thermal conductance of membranes. The sensor is constituted by a thin rectangular suspended membrane with a highly sensitive thermometer lithographied in the center of the membrane. The thermal gradient is established between the center of the membrane and the frame which is regulated in temperature. The thermal conductance is deduced from the voltage signal measured at the third harmonic appearing across the transducer.

\section{Experimental}

The experiments have been performed using a mix of the 3$\omega$ and V$\ddot{o}$lklein methods \cite{Volk,Cahill}. The measurement system consists of a heater-thermometer centered along the long axis of a rectangular membrane, which can be easily downscaled. The principle of the method consists in creating a sinusoidal Joule heating generated by an a.c. electric current flowing across a transducer. The center of the membrane is thermally isolated from the frame and hence its temperature is free to increase. The amplitude of the temperature increase or its dependence on the frequency of the excitation is entirely related to the thermal properties of the membrane. By measuring the $V_{3\omega}$ voltage appearing across the transducer, it is possible to deduce the thermal conductivity and the specific heat. The transducer is made out of a material whose resistance is strongly temperature dependant. It serves as a thermometer and heater at the same time.

\begin{figure}
\begin{center}
 \includegraphics[width=9cm]{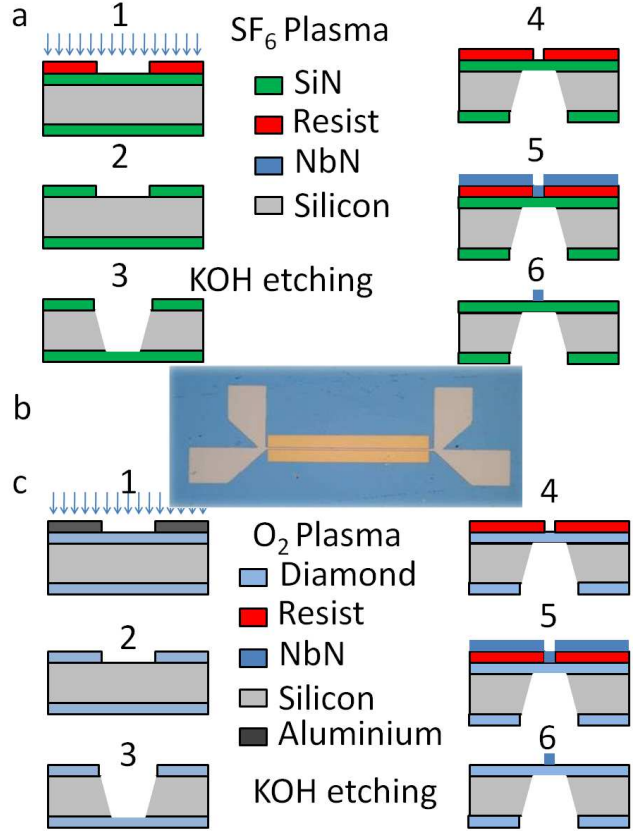}
 \end{center}
 \caption{(color on line)Schematic of the SiN (a) and diamond (c) membranes fabrication. (a) 1) The patterns of the membranes are created by photolithography. The non protected SiN is removed by SF$_{6}$ RIE. 3) The silicon is anisotropically etched in a KOH solution. 4) The thermometers are patterned by photolithography. 5) NbN is deposited by reactive sputtering. 6) The resist is removed. b) In the center of the figure a photograph of the sample is shown: the NbN is grey and the membrane is yellow(1mm long and 150$\mu$m wide).} 
 \label{fig:1}
\end{figure}

The elaboration of the membranes is detailed on the Fig. \ref{fig:1}. The amorphous SiN and polycrystalline diamond films, which have the advantage to be KOH resistant, are grown on both side of a silicon substrate by low-pressure chemical vapor deposition (LPCVD) and micro wave chemical vapor deposition (MWCVD) respectively. The 1 mm long and 150~$\mu$m large membranes are patterned on the rear side by photolithography. After removing the silicon nitride by SF$_{6}$ Reactive Ion Etching, the exposed silicon on the rear side is removed in KOH. Finally, rectangular SiN membranes are obtained on the front side. The process is similar for the diamond membrane. The rear windows are opened using photolithography. A 100~nm aluminium film protects the diamond outside the patterns. The non protected diamond is removed by O$_{2}$ Reactive Ion Etching and then the silicon is removed by KOH etching.

The transducers are patterned on the membranes by regular photolithography. They consist of niobium nitride and are grown using a dc-pulsed magnetron sputtering from a high purity (99\%-95\%) Nb target in a mixture of Ar/N$_{2}$. This type of high sensitivity thermometer is described in details elsewhere \cite{Bourgeois}. Its temperature coefficient of resistance (TCR) can be tailored over a wide temperature range, from low temperature \cite{Heron} to high temperature \cite{Lopeandia}. Hence, depending on the stoichiometry, the electrical properties of the NbN can vary a lot. For the SiN measurement, the thermometer has been designed for the 100~K-320~K temperature range and for the diamond membrane for the 10K-100K. Typically, the resistance of the thermometer is about 130~k$\Omega$ at room temperature with a TCR of 10$^{-2}$ K$^{-1}$ and 1MOhm at 70K with a TCR of 0.1K$^{-1}$. The resistance of the thermometer on membrane is calibrated using a standard four probe technique between 4~K and 330~K in a $^{4}$He cryostat.

\begin{figure}
\begin{center}
 \includegraphics[width=9cm]{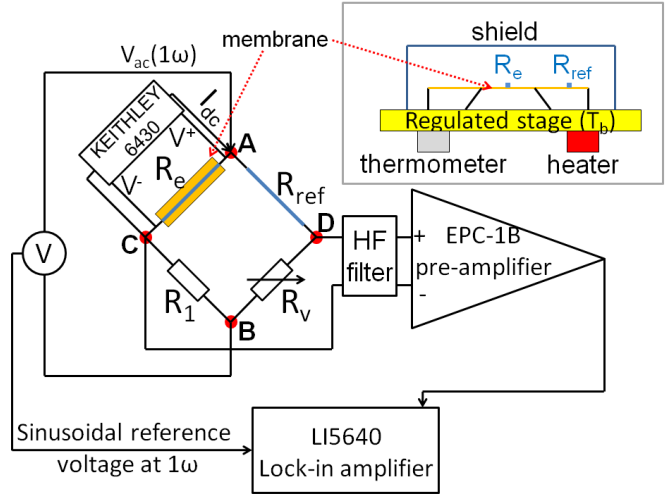}
 \end{center} 
 \caption{(color on line) Schematic of the electrical measurement system including HF filter, preamplifier and lock-in amplifier. A, B,C and D represent the nodes of the Wheatstone bridge. The $V_{3\omega}$ is measured between C and D. The transducer is referred as R$_e$ and the reference resistance as R$_{ref}$. The inset presents a schematic of the membrane fixed on the temperature regulated stage covered by the thermal copper shield.} 
 \label{fig:2}
\end{figure}

The ac current induces temperature oscillations of the membrane with an angular frequency 2$\omega$. Consequently the thermometer resistance varies with the same angular frequency. Finally, the measured voltage, due to the thermal oscillations, varies with an angular frequency 3$\omega$. This $V_{3\omega}$ voltage depends on the geometry, the thermal conductivity and the specific heat of the membrane. However, the $V_{\omega}$ signal is still present and is much larger than the $V_{3\omega}$ signal by a factor of 10$^{3}$. In the following, we explain how by using a specific Wheatstone bridge \cite{Birge} we strongly reduce the component of the measured voltage at angular frequency 1$\omega$. The bridge consists of the measured sample (resistance $R_{e}$), which is the NbN thermometer on the SiN membrane, the reference thermometer ($R_{ref}$), an adjustable resistor $R_{v}$ and an equivalent non adjustable resistor $R_{1}=50$KOhm as schematized on the Fig. \ref{fig:2}. The reference thermometer (or reference transducer) has the same geometry and is deposited in the same run as the transducer on the membrane. The two resistors $R_{v}$ and $R_{1}$ are positioned outside the cryogenic system. If $R_{e}$ = $R_{ref}$ and $R_{1}$ = $R_{v}$, the electrical potential at angular frequency $\omega$ is the same in C and D. Consequently, there is no voltage at the angular frequency of $\omega$ between C and D. Since the reference thermometer is not on the membrane, its temperature remains at $T{_b}$ and therefore, its resistance does not change. The elevation of temperature due to self heating of the reference transducer is neglected thanks to the infinite reservoir of the bulk silicon as compared to the membrane. In that geometry, the voltage at 1f ($V_{1\omega}$) has been reduced by a factor of 10$^3$. Thus, it is possible to measure the $V_{3\omega}$ signal between C and D without the 1$\omega$ component saturating the dynamic reserve of the lock-in amplifier. The two NbN thermometers have practically the same temperature behavior as they have been deposited simultaneously on the SiN substrate. However, due to the presence of inhomogeneity in the deposition process, there is a slight difference of resistance. Thus the $R_{v}$ resistor is used to balance the bridge. Thanks to the Wheatstone bridge, the $V_{3\omega}$ signal is larger than to the $V_{1\omega}$ signal.

The heating current is generated by applying an alternative voltage $V_{ac}$ between A and B with the oscillation output of the LI 5640 lock-in amplifier. The measured voltage is frequency filtered (for f$\geq$50 kHz) and preamplified by a factor of 100 with a low noise preamplifier EPC-1B \cite{Celians}. EPC-1B is a preamplifier developed at the Institut N\'eel with an input noise around 1 nV/$\sqrt{\rm{Hz}}$ between 1 Hz and 1 kHz. In the next sections, the $V_{3\omega}$ component will be presented as a function of the frequency, the amplitude of the excitation voltage $V_{ac}$ and as a function of the temperature.

The membrane is installed on a temperature regulated stage and protected by a thermal shield as schematized on the inset of the Fig. \ref{fig:2}. The copper shield, which is maintained at a temperature close to $T_{b}$, reduces strongly the radiation heat transfer. The thermal gradient between the thermal shield and the sample has been estimated to be much less than 1K, giving a power of 1Watt per meter square exchanged between the membrane and the shield. This is equivalent to a parasitic thermal conductance of $10^{-7}$W.m$^{-1}$.K$^{-1}$. Consequently, any radiative heat transfer will be neglected in the following. The stage temperature is regulated with a stability of the order of few milliKelvin. The stage temperature $T_{b}$ may be varied from 4 K to more than 330 K.
\section{Thermal and electrical models of the system}

The membrane is represented in the Fig. \ref{fig:3}a. As there is a symmetric axis coming through the middle of the transducer, the thermal system can be modeled using half the membrane and half the heating power. For simplification, as the membrane is thin ($e\leq$400 nm), we assume that the part of the membrane just beyond the NbN transducer is heated like the thermometer. As the membrane is suspended in vacuum, we assume that the heat can only diffuse through the membrane toward the silicon substrate which is at constant temperature $T_{b}$. Thus, in first approximation, we consider a one-dimensional model. The radiative heat loss is neglected as a thermal shield is put between the sample and the calorimeter wall, as explained in the previous section. Therefore, the system can be modeled as a volume of matter with a total specific heat $C'$ and bonded to the thermal bath by the membrane with a thermal conductivity $k$. The thermal system is schematized on the Fig. \ref{fig:3} b). The total specific heat $C'$ take into account both the NbN thermometer and the part of the membrane below the transducer (cf. Fig \ref{fig:3}). $C'$ can be written as: 

\begin{figure}
\begin{center}
 \includegraphics[width=9cm]{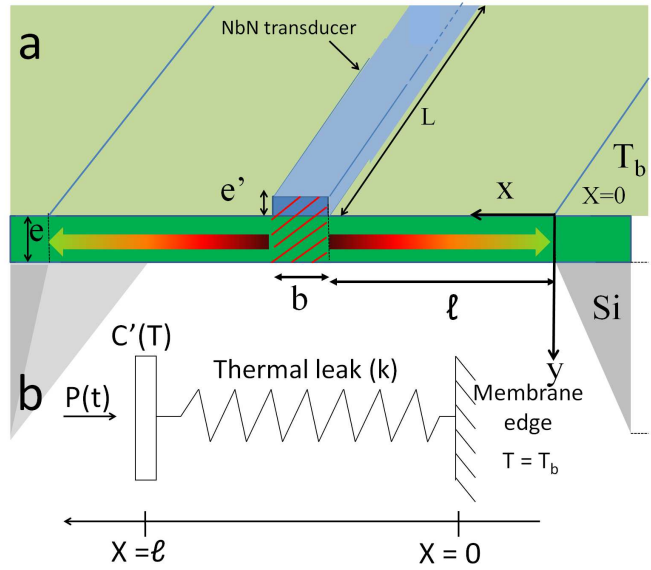}
 \end{center}
 \caption{(color on line)(a)Schematic of the membrane on which the NbN transducer has been deposited. $e$ is the thickness of the membrane and $e'$ the thickness of the NbN transducer. $C'$ is the specific heat of the hatched area in red. (b) Schematic of the thermal system.}
 \label{fig:3}
\end{figure}

\begin{equation}
C'=\rho_{NbN}c_{NbN}Le'\frac{b}{2}+c\rho\frac{b}{2}Le
\end{equation}
with $c$ the specific heat and $\rho$ the density of the SiN membrane.
The temperature is given by the 1D heat diffusion equation:
\begin{equation}
\frac{\partial^{2}T(x,t)}{\partial x^{2}}=\frac{1}{D}\frac{\partial T(x,t)}{\partial t}
\label{eqgen}
\end{equation}
with $D$ the diffusivity of the membrane. Calculations using a 2D model give approximately the same results. In order to calculate the solution of Eq. \ref{eqgen}, we need initial and boundary conditions. Therefore, we assume that at $t$=0, the temperature of the membrane is $T_{b}$ since the transducer is not heated. Moreover, we assume that the membrane edge is always at $T$=$T_{b}$. Thus, the initial and boundary conditions can be written as:
$T(x,t=0)=T_{b}$, $T(x=0,t)=T_{b}$. Moreover, the total dissipated power $P(t)$ is used to heat both the thermometer and the part of the membrane under the thermometer, and the rest of the membrane: $C'(T)\frac{\partial T(x,t)}{\partial t}_{x=\ell}$=P(t)-eLk$\frac{\partial T}{\partial x}_{x=\ell}$.

The general solution of Eq. \ref{eqgen} is:

\begin{equation}
T(x,t)=\frac{P_{0}sh\left[\omega'(1+j)x\right]e^{j2\omega t}}{(1+j)Sk\omega'ch\left[\omega'(1+j)\ell\right]+j2C'\omega sh\left[w'(1+j)\ell\right]}
\label{Tac}
\end{equation}

with $\omega'=\sqrt{\frac{\omega}{D}}$, $S$=$eL$, $P_{0}$=$\frac{RI_{0}^{2}}{4}$ and $I=I_{0}sin(\omega t)$.
We can also write Eq. \ref{Tac} using exponential notation:

\begin{equation}
T(x,t)=\frac{P_{0}}{D_{0}^{1/2}}\left[sin^{2}(\omega'x)+sh^{2}(\omega'x)\right]^{1/2}e^{j2\omega t+\varphi}
\label{tx}
\end{equation}

with $\varphi$ the phase and $D_{0}^{1/2}$ the module of the denominator of Eq. \ref{Tac}.
After development in Taylor expansion in first order in $\omega$, the expression of the temperature module $T_{m}(\ell)$ can be written as followed:

\begin{equation}
T_{m}(\ell)=\frac{P_{0}}{K_{p}{\left[1+\omega^{2}\left(4\tau^{2}+\frac{2\ell^{4}}{3D^{2}}+\frac{4\tau \ell^{2}}{3D}\right)\right]}^{1/2}}
\end{equation}
with $K_{p}$=$\frac{kS}{\ell}$, $\tau$=$\frac{C'}{K_{p}}$ and $D$ the thermal diffusivity.

\begin{figure}
\begin{center}
 \includegraphics[width=9cm]{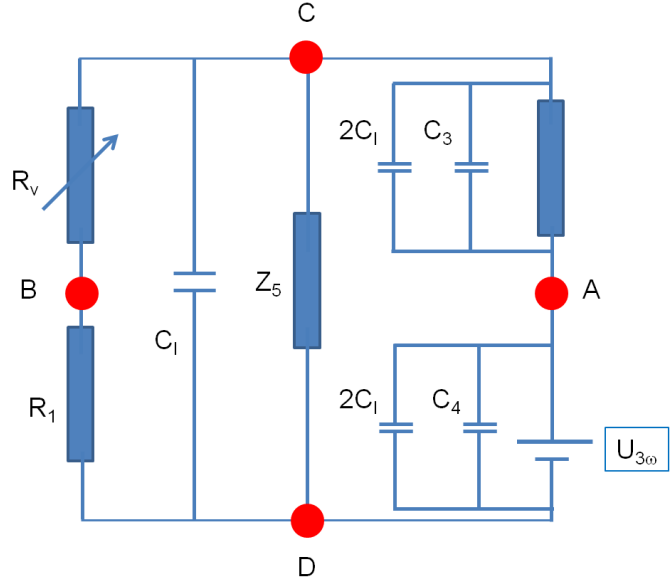}
 \end{center}
 \caption{(color on line) Electrical schematic of the Wheatstone bridge. Z$_{5}$ is the impedance of the lock-in amplifier input. The transducer is modeled as generating a voltage $U_{3\omega}$ at 3$\omega$.}
 \label{Wheat}
\end{figure}

Once the temperature variation is known, we can calculate the $V_{3\omega}$ voltage between C and D. As the distance between the thermometer and the resistances $R_{v}$ and $R_{1}$ is not negligible, we assume that there is line electric capacities ($C_{l}$) in parallel, as schematized on the Fig. \ref{Wheat}. Moreover, we assume that the two thermometers present also an electrical capacity: $C^{'}_{3}=2C_l+C_3$ for the reference and $C^{'}_{4}=2C_l+C_4$ for the sample. Besides, we consider the transducer on membrane as a 3$\omega$ voltage generator ($U_{3\omega}$) since it delivers a current at angular frequency 3$\omega$ (see Fig \ref{Wheat}). In addition, as the output impedance of the lock-in amplifier is high, we suppose that the 3$\omega$ current remains in the Wheatstone bridge system. Following the schemes given in Fig. \ref{Wheat}, the $V_{1\omega}$ and $V_{3\omega}$ voltages can be written as follows:
the module of $V_{1\omega}$:
\begin{equation}
\left|V_{1\omega}(\omega)\right|=\frac{V_{ac}\left[\epsilon^2+R^{2}_{e}R^{2}_{ref}\omega^2(R_1C^{'}_{4}-R_vC^{'}_{3})^2\right]^{1/2}}{\left[(R_v+R_e)^2+(R_vR_eC^{'}_{3}\omega)^2\right]^{1/2}\left[(R_{ref}+R_1)^2+(R_1R_{ref}C^{'}_{4}\omega)^2\right]^{1/2}}
\label{V1f(f)}
\end{equation}
with $\epsilon=R_{ref}R_1-R_eR_v$ and the phase $\varphi$ given by $\varphi=\varphi_3-\varphi_4-\varphi_5$ where :
\begin{equation}
tg\varphi_3=\frac{R_{ref}R_e(R_1C^{'}_{4}-R_vC^{'}_{3})\omega}{R_1R_{ref}-R_eR_v} \hspace{1cm} tg\varphi_4=\frac{R_vR_{ref}C^{'}_{3}\omega}{R_{ref}+R_v}  \hspace{1cm} tg\varphi_5=\frac{R_1R_eC^{'}_{4}\omega}{R_e+R_1}
\label{phas1f}
\end{equation}
and the thermal voltage generated $U_{3\omega}$ (see Fig \ref{Wheat}) is given by:
\begin{equation}
U_{3\omega}(\omega)=Z_{e}\alpha T(\ell,t)I
\end{equation}
with $Z_e=R_{e}/(1+jR_{e}C^{'}_{4}\omega$) and $I$ the current coming through the impedance $Z_{e}$.

Using the relations between the currents in the Wheatstone bridge, we can obtain the following equation:

\begin{equation}
I=\frac{V_{ac}\left(1+jR_{e}C^{'}_{4}\omega\right)}{\left(R_{1}+R_{e}+jR_{e}R_{1}C^{'}_{4}\omega\right)}
\end{equation}

Then, the general expression of $V_{3\omega}$, between C and D, can be written as follows:

\begin{equation}
V_{3\omega}(\omega)=\frac{R_{e}T_{m}e^{j\varphi}V_{ac}\alpha\left(R_{1}+R_{v}\right)\left(1+jR_{ref}C_{3}\omega\right)}{\left(R_{1}+R_{e}+jR_{1}R_{e}C^{'}_{4}\omega\right)\left[R_{1}+R_{v}+R_{ref}+jR_{ref}\left(R_{1}+R_{v}\right)\left(2C_{l}+C_{3}\right)\omega\right]}
\end{equation}

with $V_{ac}$ the voltage put on the Wheatstone bridge (between A and B),$\varphi$ the thermal phase, $C_{l}$ the line capacity (cf. Fig.), $C_{3}$ and $C_4$ being the electric capacity of the reference thermometer and of the thermometer on the membrane respectively.

The module of $V_{3\omega}$ is given by:

\begin{equation}
\left|V_{3\omega}^{rms}(\omega)\right|=\frac{R_{e}T_{m}V_{ac}^{rms}\alpha\left(R_{1}+R_{v}\right)\left[1+\left(R_{ref}C_{3}\omega\right)^{2}\right]^{1/2}}{\left[\left(R_{1}+R_{e}\right)^{2}+\left(R_{1}R_{e}C^{'}_{4}\omega\right)^{2}\right]^{1/2}\left\{\left(R_{1}+R_{v}+R_{ref}\right)^{2}+\left[R_{ref}\left(R_{1}+R_{v}\right)\left(2C_{l}+C_{3}\right)\omega\right]^{2}\right\}^{1/2}}
\label{module3w}
\end{equation}

and the phase by:

\begin{equation}
\varphi_{V_{3\omega}}(\omega)=\varphi+arctan\left(R_{ref}C_{3}\omega\right)-arctan\left(\frac{R_{e}R_{1}C^{'}_{4}\omega}{R_{1}+R_{e}}\right)-arctan\left[\frac{R_{ref}\omega\left(R_{1}+R_{v}\right)\left(2C_{l}+C_{3}\right)}{R_{1}+R_{v}+R_{ref}}\right]
\label{phase3w}
\end{equation} 

All the fits on $V_{1\omega}$  and $V_{3\omega}$ of this work have been realized with these equations.
The Fits of our results show that, for frequency below 1 kHz, the electric capacities of the thermometers are negligible as compared to the thermal effects. Thus, in order to give a clearer and more physical description of the method, we can consider that $C_{3}$=$C^{'}_{4}$=$C_{l}$=0. Then the expression of $V_{3\omega}$ becomes:

\begin{equation}
\left|V_{3\omega}^{rms}(\omega)\right|=\frac{\alpha   (V_{ac}^{rms})^{3}(R_{1}+R_v)R_{e}^{2}}{2K_{p}\left(R_{1}+R_{e}\right)^{3}\left(R_{1}+R_{e}+R_v\right)\left[1+\omega^{2}\left(4\tau^{2}+\frac{2l^{4}}{3D^{2}}+\frac{4\tau l^{2}}{3D}\right)\right]^{1/2}}
\label{V3w} 
\end{equation}

At low frequency, the $\omega$ term becomes negligible and the expression of $V_{3\omega}$ can be written in its simpler form as:
\begin{equation}
\left|V_{3\omega}^{rms}(\omega)\right|=\frac{\alpha (V_{ac}^{rms})^{3}(R_{1}+R_v)R_{e}^{2}}{2K_{p}\left(R_{1}+R_{e}\right)^{3}\left(R_{1}+R_v+R_{ref}\right)}
\label{V3wlf}
\end{equation}
Thus, at low frequency, $V_{3\omega}$ is constant and depends on the thermal conductivity $k$ of the membrane. After the calibration of $R_e$ and $R_{ref}$,$R_{1}$ and $R_{v}$ being fixed, a measure of $V_{3\omega}$ at low frequency allows the calculation of the thermal conductance $K_p$ using Eq. \ref{V3wlf}. 


\begin{figure}
\begin{center}
 \includegraphics[width=9cm]{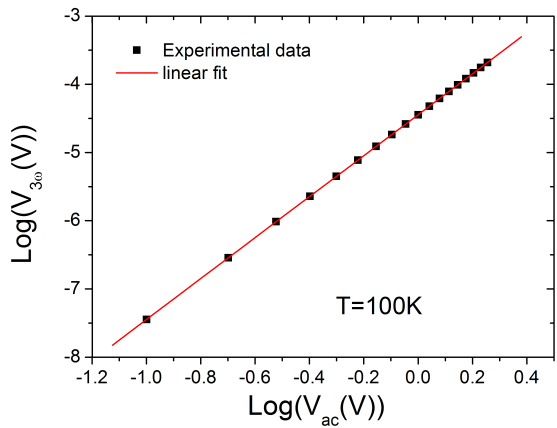}
 \end{center}
 \caption{(color on line)$V_{3\omega}$ signal as a function of the voltage applied across the Wheatstone bridge ($V_{ac}$) in logarithmic scales.}
 \label{V3wV}
\end{figure}

\section{Results}

The method has been checked using two different measurements. According to equation \ref{V3wlf}, the $V_{3\omega}$ signal depends on the cube of the Wheatstone bridge voltage $V_{ac}$. Thus, the $V_{3\omega}$ signal has been measured at different temperatures to check this behaviour. As an example, the 100~K measurement can be seen on the Fig.\ref{V3wV}. The linear fit gives a slope very close to 3 which confirms the cubic behaviour of the $V_{3\omega}$ signal versus the applied voltage $V_{ac}$.

\begin{figure}
\begin{center}
 \includegraphics[width=9cm]{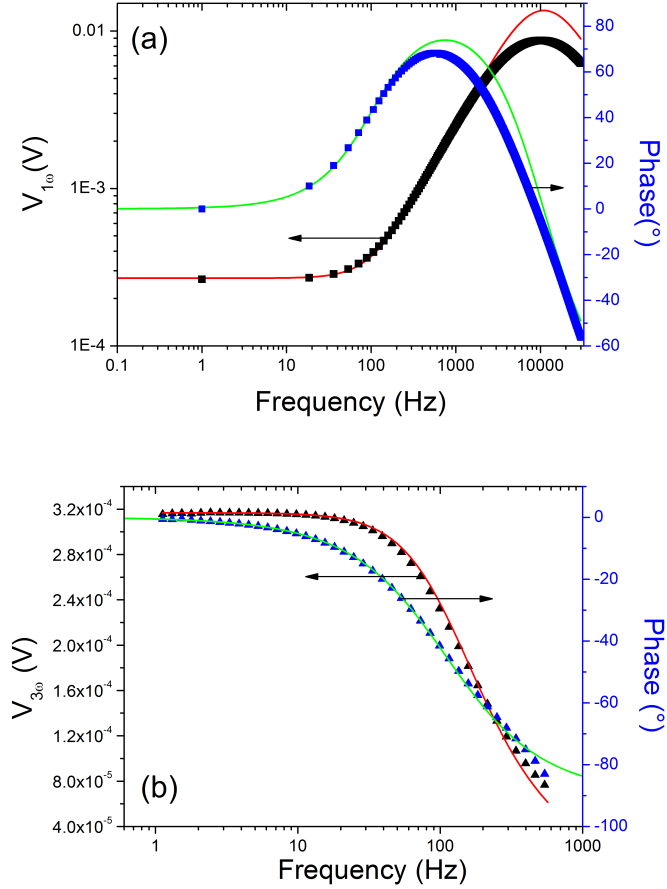}
 \end{center}
 \caption{$V_{1\omega}$ (a) and $V_{3\omega}$ (b) module and phase signals as a function of the frequency and their respecting fits (solid lines) using the following parameters in the equations \ref{Tac}, \ref{V1f(f)}, \ref{phas1f}, \ref{module3w} and  \ref{phase3w}: $V^{rms}_{ac}$=0.06V, $R_v$=70.5K$\Omega$ ,$R_e$=267.81K$\Omega$, $R_{ref}$=390.77kK$\Omega$, $C^{'}_{3}$=185pF, $C^{'}_{4}$=450pF.}
 \label{V3wf}
\end{figure}

The second way of verification needs a frequency scan measurement. According to the equation \ref{V3w}, the $V_{3\omega}$ signal depends strongly on the angular frequency $\omega$. At low frequency the square root term tends to 1 and consequently, the $V_{3\omega}$ signal becomes constant. An example of frequency scan of the $1f$ and $3f$ component are presented at the Fig. \ref{V3wf}. As predicted by the theory, a plateau is observed at low frequency for the $V_{3\omega}$ signal. The thermal conductivity $k$ can be extracted from the low frequency plateau thanks to the equation \ref{V3wlf}. Concerning the $V_{1\omega}$ signal, an electrical cutoff frequency is observed at 1KHz correctly fitted by the equations \ref{V1f(f)} and \ref{phas1f}. The thermal cutoff frequency (around 100Hz much lower than the electrical cutoff frequency) is not directly related to the characteristic time $K/C$ of the membrane due to possible thermal resistances between the thermometer and the membrane; this will be the object of further studies \cite{comment}.

\subsection{Noise, sensitivity and addenda}

\begin{figure}
\begin{center}
 \includegraphics[width=9cm]{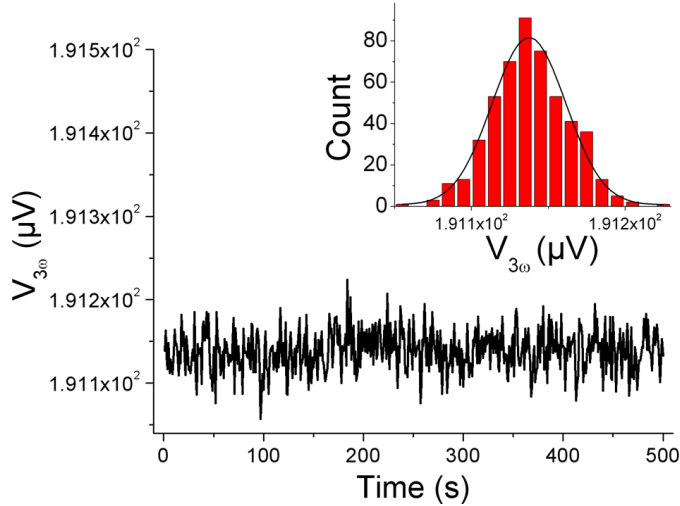}
 \end{center}
 \caption{Measure of the $V_{3\omega}$ voltage as a function of time. The inset shows the signal distribution. The full width at half maximum corresponds to the noise of the measurement. In this case, the noise is evaluated to be around 100~nV/$\sqrt{\rm{Hz}}$.}
 \label{noise}
\end{figure}

For high sensitivity measurement systems, low noise measurement systems are required. The noise has been evaluated measuring the $V_{3\omega}$ signal during 500 seconds. The measure at 120 K is presented on the figure \ref{noise}. According to this measure, the noise is about 100 nV/$\sqrt{\rm{Hz}}$ which is the expected Johnson noise. The noise/signal ratio (resolution) is about 2.10$^{-3}$. Thus the smaller thermal conductance that can be measured is below 10$^{-8}$ W/K, close to the nanoWatt per Kelvin. These measurements have been obtained with an oscillation of the temperature of the membrane of 10 to 100mK, so it means that the set-up can measure energies far below the nanoWatt. The accuracy of the method is estimated to be less than one percent.

The contribution of the thermometer to the heat conduction needs to be discussed. Indeed, the thermometer evacuates some heat to the frame through its connections on the edge of the membrane. There is actually two thermal paths to estimate, one from the electrons and one from the phonons. Using the Wiedemann-Franz law, we can give a number for the electronic thermal conductance because we know exactly the resistance of the NbN transducer. At 300K, the resistance is around 100~KOhm (and higher at lower temperature), then an electronic thermal conductance of 10$^{-10}$W/K is expected. For the phonon contribution, the NbN materials can be modeled as a glass seeing its highly amorphous structure. If we take a thermal conductivity of 10Wm$^{-1}$K$^{-1}$ and we calculate the thermal conductance taking into account the transducer geometry (thickness $e^{'}=50$nm, width $b=20\mu$m and length 1mm) we get a phonon thermal conductance of 10$^{-8}$W/K. The typical thermal conductance measured on the SiN membrane lies around few 10$^{-5}$W/K, which is much bigger than any of the two numbers given above for the electron and phonon contributions from the thermometer. Then the parasitic thermal paths due to the electrons and phonons of the transducer can be neglected.

\subsection{Thermal conductivity of SiN membrane}
\begin{figure}
\begin{center}
 \includegraphics[width=9cm]{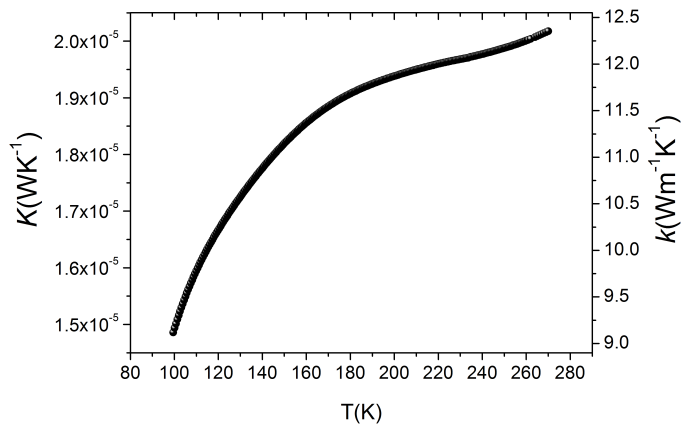}
 \end{center}
 \caption{Thermal conductivity of a 100 nm thick SiN membrane as a function of the temperature.}
 \label{kSiN}
\end{figure}

The thermal conductivity of the silicon nitride membrane has been measured between 100 and 330 K. The results for a 100 nm thick silicon nitride membrane are shown in the Fig. \ref{kSiN}.

The general behaviour is an increase of the thermal conductivity as the temperature is increased, a usual trend reported in the literature for amorphous silicon nitride films \cite{Jain,zink2009}. A value close to 10Wm$^{-1}$K$^{-1}$ is obtained at room temperature in agreement with previous measurements of thermal conductivity on thin silicon nitride membranes \cite{Jain,zink2005}.

\subsection{Thermal conductivity of diamond membrane}

\begin{figure}
\begin{center}
 \includegraphics[width=9cm]{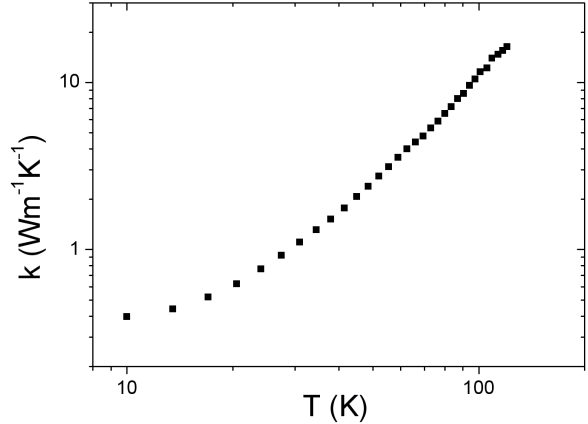}
 \end{center}
 \caption{Thermal conductivity of a 400 nm thick polycrystalline diamond membrane as a function of the temperature.}
 \label{kdiam}
\end{figure}

The thermal conductivity of a 400 nm thick polycrystalline diamond membrane has been measured between 10 and 130 K. To our knowledge, this has never been realized for a so thin diamond membrane. The results are gathered in the Fig. \ref{kdiam}. The thermal conductivity of polycrystalline diamond depends strongly on the grain size due to phonon scattering effects \cite{Angad}. The thermal conductivity ranges from 1 and 10Wm$^{-1}$K$^{-1}$ between 10 and 100K in good agreement with the literature since the data are located just between the values of microcrystalline and ultra nanocrystalline diamond films \cite{Shamsa,Graeb}.

\section{Conclusion} 
We have demonstrated a method to measure thermal conductivity of very thin suspended membranes based on a mix of the 3$\omega$ method and of the V$\ddot{o}$lklein method. The performances of the thermometry as well as the measurement chain have been optimized to obtained a very sensitive technique (down to 0.1 nanoWatt) with a resolution of approximately 10$^{-3}$. As an illustration of the technique, low stress silicon nitride membrane as well as polycrystalline diamond membrane have been measured. By adapting the thermometry, measurement from 4K to 400K can be done easily.

This method may be used to measure the in-plane thermal conductivity of various dielectric suspended materials. It is complementary to other technique more sensitive to out-of-plane thermal properties like the 3$\omega$ method applied to semi-infinite media and thermoreflectance. Moreover, SiN membrane can be used as a sample holder for the deposition of thin film on the back side of the sensor; it may serve as a platform for thermal characterization of any thin films.

We acknowledge support from Nanofab, Cryogenic shop, Electronic shop and Capthercal for these experiments. Funding for this project was provided by a grant from La R\'egion Rh\^one-Alpes (Cible and CMIRA) and by the Agence Nationale de la Recherche (ANR) through the project SENSOCARB. We would like to thank P. Gandit, P. Brosse-Maron, B. Fernandez, T. Fournier, C. Donnet, F. Garrelie and J.-L. Garden for help and fruitful scientific exchanges.


\begin{thebibliography}{99}
\bibitem{Celians}
EPC1B preamplifier has been developed at the Institut N$\acute{e}$el and commercialized by Celians-C3EM.
\bibitem{comment} This frequency dependence of the V$_{3\omega}$ signal is still under study, the real physical origin of the cutoff has not been found yet. Fortunately it is not needed for the exact extraction of the thermal conductivity value of the membrane.
\end{thebibliography}
\end{document}